# Optical Trajectory of the Insulator-to-metal Transition of Ultra-thin Gold Films

**Xuefeng Wang, Ming Zhao and D.D. Nolte**
*Department of Physics, Purdue University, West Lafayette, IN 47907*
nolte@physics.purdue.edu

We apply interferometric picometrology to measure the complex refractive index and dielectric properties of ultra-thin gold films as a continuous function of thickness from 0.2 nm to 10 nm deposited on thermal oxide on silicon. Three distinct regimes are found for the dielectric constant trajectory in the complex plane. The first regime shows a circular trajectory predicted by the Drude equation tracing the gold bulk-to-thin film transition. The second and third regimes are manifested as metal clusters range from large to small size and sparse coverage. The entire atom-to-bulk transition of gold is revealed from the optical perspective.

PACS numbers:

A thin metal film has different optical properties than the bulk [1, 2]. However, there has been no study revealing the entire atom-to-film-to-bulk transition of the complex refractive index $\tilde{n}$ and dielectric function $\tilde{\varepsilon}$ when the metal film thickness is below 1 nm. The ultra-thin regime is limited by current techniques for thin film studies. The complex function $\tilde{n}$ of a thin film has required multi-parameter measurements (wavelength spectrum, polarization or angle dependence) [3]. Thin films are usually deposited on a substrate because the substrate provides a flat and supporting platform for the films. The film modifies the reflectivity $\tilde{r}$ of the substrate, and the refractive index of a film is concealed in the change in $\tilde{r}$. Reflectometry measures the modulus change $|\tilde{r}|$ of the reflection coefficient due to the film, but this one-parameter measurement cannot acquire the full information needed for the complex refractive index calculation. As a compromise, the reflectance change may be measured in a broad wavelength range and the Kramers-Kronig relation is applied to find $\tilde{n}$ [4, 5]. This can introduce errors due to insufficient wavelength bandwidth and makes detection complicated and sometimes inapplicable for small sample size. In contrast to changing the wavelength, ellipsometry measures $\tilde{n}$ by monitoring reflectances at two different polarization states of oblique-incident light [6, 7]. However the basic model of ellipsometry requires that $\tilde{n}$ is equal at different polarization states, and it becomes difficult or inapplicable if the films have optical anisotropy which usually occurs to the sub-nanometer metal film . Ellipsometric measurement becomes erratic when the gold film thickness is below 5.8 nm [8].

In this paper, we use interferometric picometrology to study $\tilde{n}$ and $\tilde{\varepsilon}$ through the atom-to-film-to-bulk transition of gold for thicknesses ranging continuously from 0.1 nm to 10 nm at normal incidence and at a single wavelength. The complex change of the $\tilde{r}$ of the substrate due to the film is acquired to calculate $\tilde{n}$ of the film by analyzing the far-field diffraction of a gold-patterned surface. The modulus change of $|\tilde{r}|$ is obtained by monitoring the intensity change of the diffraction pattern, while the phase change $\varphi(\tilde{r})$ is obtained by monitoring the spatial shift of the diffraction pattern caused by the asymmetric diffraction of the reflected beam when the focal spot scans over the edge of a thin film. In the experiments, both the modulus and the phase change of $\tilde{r}$ caused by the thin film are acquired using a quadrant detector that simultaneously monitors both intensities and spatial shifts of the reflected beam [9, 10]. Once the output intensity (I) and phase contrast (PC) signals are measured, $\tilde{n}$ is calculated using Eq.1 after the normalized amplitudes of I and PC signals are acquired

$$(\tilde{n}^2 - 1)\frac{(1+\tilde{r})^2}{\tilde{r}}\frac{2\pi d}{\lambda} = 3.565 A\left[i^{PC}(x)\right] + A\left[i^{I}(x)\right] j \qquad (1.)$$

where $\tilde{r}$ is the reflection coefficient of the bare substrate. the thickness of the film is $d$, $\lambda$ is the wavelength in vaccum, $i^I(x)$ and $i^{PC}(x)$ are the normalized signals of the I and PC channels, and $A[i^I(x)]$ and $A[i^{PC}(x)]$ represent the amplitudes of $i^I(x)$ and $i^{PC}(x)$.

Picometrology is based on Gaussian beam diffraction study from the edge of a film [9, 11]. Therefore, we deposited gold films in a stripe pattern (Fig.1 a) using photolithography to create a periodic pattern on a silicon chip with a nominal 140 nm thermal oxide layer (measured precisely to be 134 nm by a spectroscopic analysis of the reflectance). The periodic pattern along the x-direction provides multiple edges of films and improves the detection sensitivity. In the y-direction, the gold film thickness increases continuously and linearly from 0 nm to 10 nm. This is achieved by a metal evaporator modified by attaching a step-motor-controlled plate between the evaporation source and the silicon chip. The plate moves slowly and exposes the chip surface continuously and incrementally during evaporation. With this method, a stripe-patterned gold sample is created, and the optical study from 0 to 10 nm is performed on a single chip. The error due to variations among different chips is eliminated. The gold thickness at any location can be determined by finding the start (0 nm) and end deposition (10 nm) position. In experiments, the gold deposition rate is 0.2 nm/s, and the film is not annealed.

We performed measurements on the gold sample at the wavelengths 488 nm and 532 nm with a scanning system shown in Fig. 1. Two-dimensional scanning is realized by a platform consisting of a spinner and a linear stage. The reflected beam is analyzed by projecting the beam on a split detector located on the Fourier plane. The detector consists of two semi-circlular halves A and B. The intensity (I) signal of the diffraction pattern is acquired as I = A+B, and the phase contrast (PC) signal is acquired PC = A-B. Images of I and PC scans are shown in Fig. 2b and 2c. The normalized amplitudes of the I and PC signals are found for any gold thickness by extracting single lines of data from the I and PC images at the corresponding positions. Normalization is done by dividing the I and PC signal by the intensity of the reflected beam on the land of the substrate. As a demonstration, in Fig. 3a, the I and PC signals are extracted for 4 nm gold and $\tilde{n}$ of the gold can be calculated at this thickness. Although the approximate Eq. 1 is suitable to calculate $\tilde{n}$ when the thickness is much less than probe wavelength, for the highest thickness of 10 nm we use a rigorous computational transfer matrix method to find the precise $\tilde{n}$. The $\tilde{r}$ of the substrate, the film thickness $d$, and I and PC responses are known, hence we plot $\tilde{n} = n - ik$ ($0 \leq n, k \leq 4$, 0.01 as for each step) on the complex plane to find the $\tilde{n}$ satisfying both the I and PC responses. Calculations show that there is a single curve on which all $\tilde{n}$ give and equal I response (which explains why reflectometry alone cannot determine the $\tilde{n}$). All $\tilde{n}$ satisfying the PC response form a different curve. The intersection of the I and PC curves uniquely determines the target $\tilde{n}$ (Fig.3 b). By these means, the $\tilde{n}$ of gold is calculated for any thickness less than 10 nm. The curves of $\tilde{n} = n - ik$ are shown in Fig.3 d.

The trajectory of the dielectric constant $\varepsilon = \tilde{n}^2$ across the complex plane, parametrized by the film thickness, is shown in Fig. 4. Three distinct regimes are observed in the $\varepsilon$ trajectories, and these regimes are manifested in three different topologies of gold on the substrate. The first regime (gold thickness: bulk ~ 2 nm) reveals an evolution from thick film to thin film. In this regime, the gold film is continuous or at least consists of laterally large clusters. The mean path of free electrons (25 nm in Au [12]) is shortened only due to thin film thickness. The collision frequency of free electrons increases and changes the dielectric constant of gold. The Drude equation (Eq. 2) can be used to describe $\varepsilon$ of gold for this topology. The dielectric constant exhibits a circular trajectory on the complex plane if $\Gamma$ changes at constant $\omega$ (Eq. 3). The radius of this circle is $\omega_p^2/2\omega^2$ and the center is $\varepsilon_{bound} - \dfrac{\omega_p^2}{2\omega^2}$ where $\omega_p$ is plasmon frequency, $\omega$ is the probe light requency and $\varepsilon_{bound}$ is the dielectric constant contributed by bound electrons. The eqations describing these radii are

$$\varepsilon = \varepsilon_{bound} - \frac{\omega_p^2}{\omega^2 + i\Gamma\omega} \qquad \text{where } \Gamma = \Gamma(bulk) + kv_F/H \tag{2.}$$

$$\left|\varepsilon - (\varepsilon_{bound} - \frac{\omega_p^2}{2\omega^2})\right| = \left|\left(\frac{-1+i\Gamma/\omega}{1+i\Gamma/\omega}\right)\frac{\omega_p^2}{2\omega^2}\right| = \frac{\omega_p^2}{2\omega^2} \tag{3.}$$

and from the data on Fig. 4 the radii are 3.8 at 488 nm and 5.2 at 532 nm.

In the second regime, a transition of $\varepsilon$ to a linear behavior is observed, possibly because the lateral sizes of partial clusters become smaller than the mean free path. In this regime the lateral size of gold clusters contributes to the change of $\Gamma$. The topology can be treated as a mixture of large and small clusters (the mean free path is the threshold for the sense of "large" and "small") with a ratio $f_s$ which is the percentage of small clusters covering the surface (from 0 to 1). We use the Maxwell-Garnett equation [13-15]

$$\frac{\varepsilon - \varepsilon_A}{L\varepsilon + (1-L)\varepsilon_A} = f_B \frac{\varepsilon_B - \varepsilon_A}{L\varepsilon_B + (1-L)\varepsilon_A} \tag{4.}$$

to interpret the results, where the effective dielectric constant $\varepsilon$ is expressed in terms of $\varepsilon_B$ as inclusions (gold) inside the medium $\varepsilon_B$ (air). The volume fraction of the inclusions $f_b$ is the fraction of material b in medium a. The depolarization factor L equals 1/3 for spherical clusters or 0 for flat metallic plates whose normals are perpendicular to the electric field [13]. In the case of large+small clusters, L approximately equals 0 because the lateral size is much larger than the height of clusters in this regime. In this case Eq. 4 is reduced to Eq. 5

$$\varepsilon = f_s(\varepsilon_s - \varepsilon_l) + \varepsilon_l \tag{5.}$$

which is a linear function of $f_s$ (volume fraction of small clusters), and $\varepsilon_s$ and $\varepsilon_l$ are the dielectric constants of the small and large clusters, respectively. This matches the observed linear transition of $\varepsilon$ to the second regime. By extrapolation, the $\tilde{n}$ of films consisting of extremely fine clusters, or gold atoms can be calculated to be 2.6 at 488 nm and 3.2 at 532 nm.

In the third and thinnest regime, average gold thickness is smaller than 1 nm, and the gold clusters become small and sparse. Clusters cannot fully cover the surface, and the topology of the gold film consists of gold clusters and air. The clusters are treated as spheres with L = 1/3 in Eq. 4. We calculate the effective $\varepsilon$ of the mixture of air and gold clusters with Eq. 6, where $f_s$ is the volume fraction of small clusters (from 1 to 0). The real part of $\varepsilon$ changes from 6.0 to 3.0 and the imaginary part changes from 1 to 0 as predicted by Eq. 4, derived from the Maxwell-Garnett rule

$$\varepsilon = \frac{3(\varepsilon_s - 1)}{3 + (1 - f_s)(\varepsilon_s - 1)} + 1 \tag{6.}$$

Therefore, the three distinct behavior regimes that appear in Fig. 4 may be explained through effective dielectric constants that move through three successive topology regimes. As a function of increasing deposition, isolated gold atoms combine to form small clusters, then evolve into a composite of small clusters on large flat rafts of gold film, followed by a more homogeneous layer of gold, until the bulk values of gold are approached for thickness above 10 nm. The breakpoints between these regimes are 0.6 nm and 2 nm.

AFM and SEM images of gold films from 1 nm to 10 nm are presented in Ref. [16]. where similar deposition conditions were applied to deposit gold thin films on a glass surface. The topologies of gold match well with our results.

In conclusion, we have measured the trajectory in the complex plane of the dielectric constants of thermally-evaporated gold on silica at the wavelengths 532 nm and 488 nm as a function of gold film thickness. Picometrology is able to extract the full complex dielectric constant using a single-wavelength measurement without polarization control, and at normal incidence, significantly simplifying the measurements and enabling experiments in the ultrathin limit (0.1 nm thickness). Most importantly, the experimentally-extracted values for $\tilde{n}$ and $\tilde{\varepsilon}$ are model-independent (in constrast to ellipsometry), although the explanation of the trajectory across the complex plane does invoke specific topology. The physical behavior with increasing gold deposition goes through three topological regimes that involve the clustering of gold.

**Reference:**

[1]   R. B. Laibowitz and Y. Gefen, "Dynamic Scaling near the Percolation-Threshold in Thin Au Films," *Physical Review Letters*, vol. 53, pp. 380-383, 1984.
[2]   J. J. Tu, C. C. Homes, and M. Strongin, "Optical properties of ultrathin films: Evidence for a dielectric anomaly at the insulator-to-metal transition," *Physical Review Letters*, vol. 90, 2003.
[3]   A. Gray, M. Balooch, S. Allegret, S. De Gendt, and W. E. Wang, "Optical detection and characterization of graphene by broadband spectrophotometry," *Journal of Applied Physics*, vol. 104, 2008
[4]   P. Grosse and V. Offermann, "Analysis of Reflectance Data Using the Kramers-Kronig Relations," *Applied Physics a-Materials Science & Processing*, vol. 52, pp. 138-144, 1991.
[5]   D. A. Crandles, F. Eftekhari, R. Faust, G. S. Rao, M. Reedyk, and F. S. Razavi, "Kramers-Kronig-constrained variational dielectric fitting and the reflectance of a thin film on a substrate," *Applied Optics*, vol. 47, pp. 4205-4211, 2008.
[6]   F. L. McCrackin, E. Passaglia, R. R. Stromberg, and Steinber.Hl, "Measrumemt of Thickness and Refractive Index of Very Thin Films and Optical Properties of Surfaces by Ellipsometry," *Journal of Research of the National Bureau of Standards Section a-Physics and Chemistry*, vol. A 67, pp. 363-&, 1963.
[7]   R. J. Archer, "Determination of Properties of Films on Silicon by Method of Ellipsometry," *Journal of the Optical Society of America*, vol. 52, pp. 970-&, 1962.
[8]   M. Yamamoto and T. Namioka, "Insitu Ellipsometric Study of Optical-Properties of Ultrathin Films," *Applied Optics*, vol. 31, pp. 1612-1621, 1992.
[9]   X. Wang, M. Zhao, and D. D. Nolte, "Common-Path Interferometric Detection of Protein on the BioCD," *Appl. Opt.*, vol. 46, pp. 7836-7849, 2007.
[10]  M. Zhao, W. Cho, F. Regnier, and D. Nolte, "Differential phase-contrast BioCD biosensor," *Appl. Opt.*, vol. 46, pp. 6196-6209, 2007.
[11]  X. Wang, M. Zhao, and D. D. Nolte, "Strong Anomalous Optical Dispersion of Graphene: Complex Refractive Index Measured by Picometrology," *Opt. Express*, vol. to be published, 2009.
[12]  M. Walther, D. G. Cooke, C. Sherstan, M. Hajar, M. R. Freeman, and F. A. Hegmann, "Terahertz conductivity of thin gold films at the metal-insulator percolation transition," *Physical Review B*, vol. 76, 2007.
[13]  R. W. Cohen, G. D. Cody, M. D. Coutts, and B. Abeles, "Optical Properties of Granular Silver and Gold Films," *Physical Review B*, vol. 8, pp. 3689-3701, 1973.
[14]  B. Abeles and J. I. Gittleman, "Composite-Material Films - Optical-Properties and Applications," *Applied Optics*, vol. 15, pp. 2328-2332, 1976.
[15]  J. I. Gittleman and B. Abeles, "Comparison of Effective Medium and Maxwell-Garnett Predictions for Dielectric-Constants of Granular Metals," *Physical Review B*, vol. 15, pp. 3273-3275, 1977.
[16]  I. Doron-Mor, Z. Barkay, N. Filip-Granit, A. Vaskevich, and I. Rubinstein, "Ultrathin gold island films on silanized glass. Morphology and optical properties," *Chemistry of Materials*, vol. 16, pp. 3476-3483, 2004.

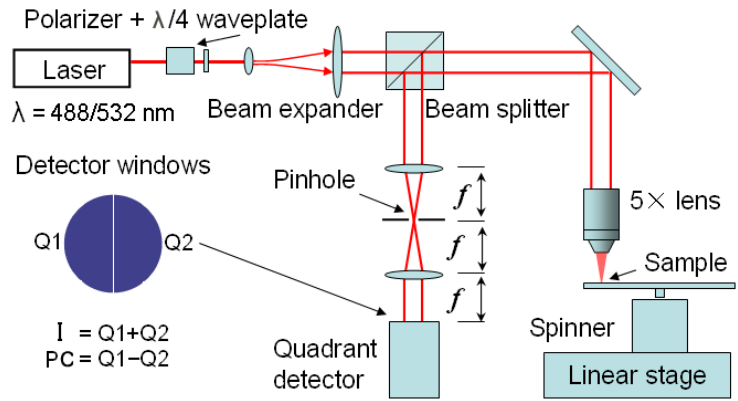

FIG. 1 (color online). Schematics of picometrology. Picometrology measures complex refractive index of ultra-thin films at single wavelengths at normal incidence. The reflected Gaussian beam forms an asymmetric diffraction pattern on the Fourier plane when scanning across film edges. By analyzing the spatial shift and intensity variation with a split detector, full information is acquired for the complex refractive index calculation.

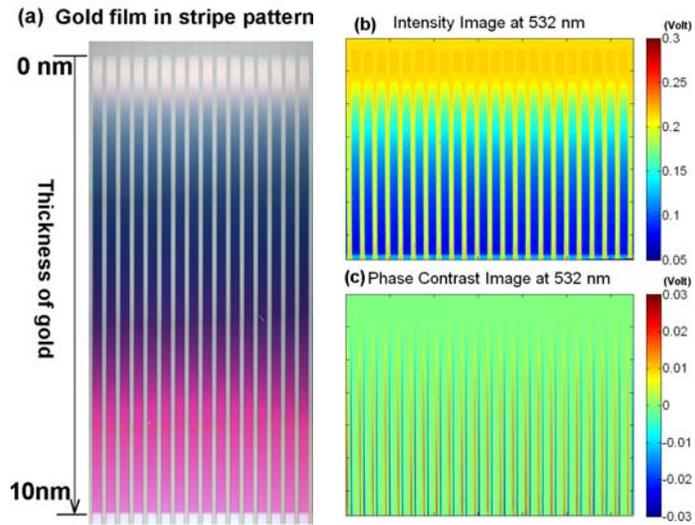

FIG. 2 (color). (a) Stripe-patterned gold film with thickness varying continuously from 0 to 10 nm on thermal oxide on silicon (140 nm $SiO_2$). The sample is prepared by a thermal metal evaporator as the evaporation time is controlled by a stepper motor. The image is captured by microscope under white light. The color shows a rich transition within 10 nm (Real color). (b) and (c) show the images of the sample in the I and PC channels scanned by the picometroloy system at 532 nm (Pseudo color).

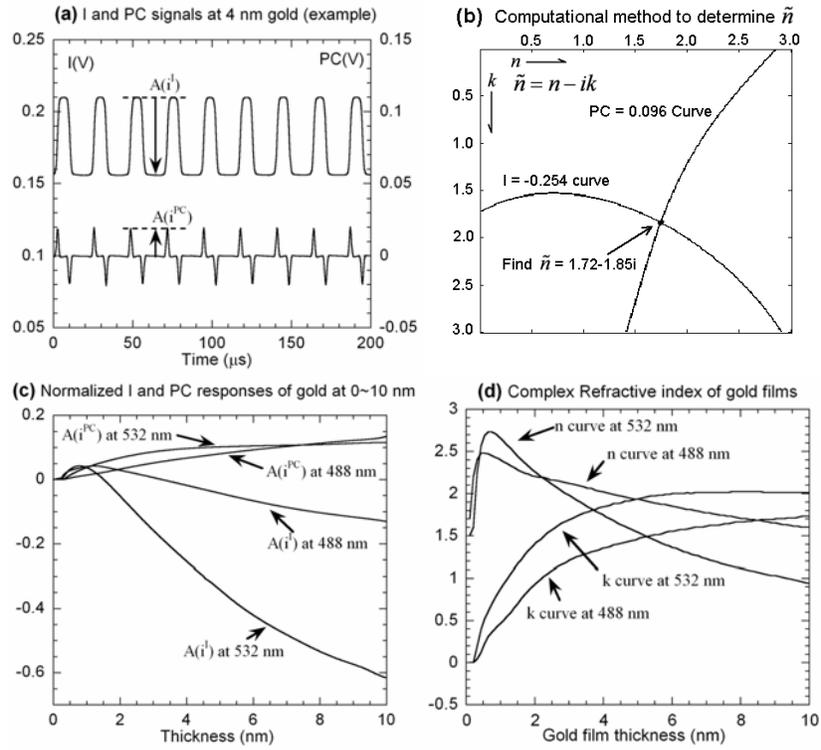

Fig. 3. (a) One example of I and PC signals at 4 nm gold thickness (one track taken from image Fig. 1b and 1c). The normalized amplitudes of both channels are acquired, and the refractive index of gold at this thickness can be calculated by Eq. 1. (b) The I response is satisfied by multiple $\tilde{n}$ which form a curve on the complex plane, and similarly for the PC response. The intersection point uniquely determines $\tilde{n}$ of the film. (c). Normalized amplitudes of I and PC responses of gold from 0 to 10 nm at 488 nm and 532 nm wavelengths. (d) The $\tilde{n}$ is calculated for gold films with 0~10 nm thicknesses at both 488 nm and 532 nm wavelengths.

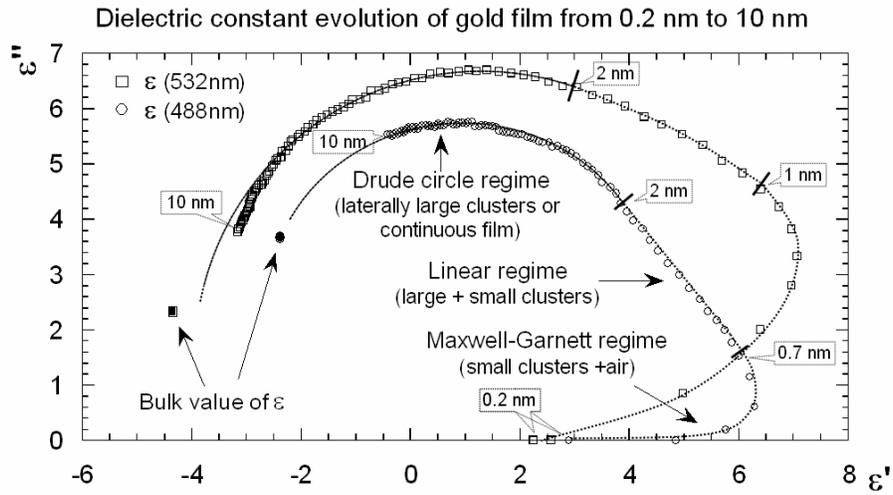

Fig. 4. The dielectric constants of gold that are derived from the complex refractive index (Fig. 1c). The dielectric trajectory has three distinct regimes that originate from three types of gold topology when the thickness changes from bulk down to 0.2 nm.